\documentclass[10pt,aps]{revtex4}
\usepackage{graphicx}
\begin{document}

\title{Influence of conformational fluctuations on enzymatic
activity: modelling the functional motion of $\beta$-secretase}

\author{M. Neri$^1$, M. Cascella$^{1,2}$ and C. Micheletti$^1$}
\address{(1) SISSA and
INFM, Via Beirut 2-4, I-34014 Trieste, Italy\\
(2) EPFL, Lausanne, Switzerland} 
\date{\today} 
\begin{abstract}
Considerable insight into the functional activity of proteins and
enzymes can be obtained by studying the low-energy conformational
distortions that the biopolymer can sustain. We carry out the
characterization of these large scale structural changes for a protein
of considerable pharmaceutical interest, the human $\beta$-secretase.
Starting from the crystallographic structure of the protein, we use
the recently introduced $\beta$-Gaussian model to identify, with
negligible computational expenditure, the most significant distortion
occurring in thermal equilibrium and the associated time scales.  The
application of this strategy allows to gain considerable insight into
the putative functional movements and, furthermore, helps to identify
a handful of key regions in the protein which have an important
mechanical influence on the enzymatic activity despite being spatially
distant from the active site. The results obtained within the Gaussian
model are validated through an extensive comparison against an
all-atom Molecular Dynamics simulation.
\end{abstract}
\maketitle

\section*{Introduction}

In the past decades several experimental and theoretical studies have
pointed out that, although proteins possess an atomic density
comparable to that of crystalline solids, they are much more flexible
that the latter\cite{levitt85,go_gauss91,bav93}. This unusual
elasticity, whose origin putatively resides in the neat secondary and
tertiary protein organization~\cite{normod}, is a prerequisite for
biological functionality. In fact, in order to carry out their
biological tasks, proteins and enzymes need to sustain conformational
distortions where groups of several amino acids are significantly
displaced from the reference native states. The time scales associated
to such configurational changes, which may occur over several
nanoseconds, is also long compared to those of atomic
motions~\cite{revTru}.

From a computational point of view, the most direct way of observing
these rearrangements would be to resort to a Molecular Dynamics
simulation~\cite{karplus_ACR}. Present implementations of this scheme
allow to follow the dynamical evolution of a large protein (of a few
hundred residues) in its surrounding solvent for about one nanosecond.
This time scales is large enough to gain considerable insight into
several dynamical aspects of large proteins but may be inadequate to
characterize accurately the functional movements mentioned
before~\cite{hessb}.

\begin{figure}[t]
\begin{center}
(a)\includegraphics[width=2.5in]{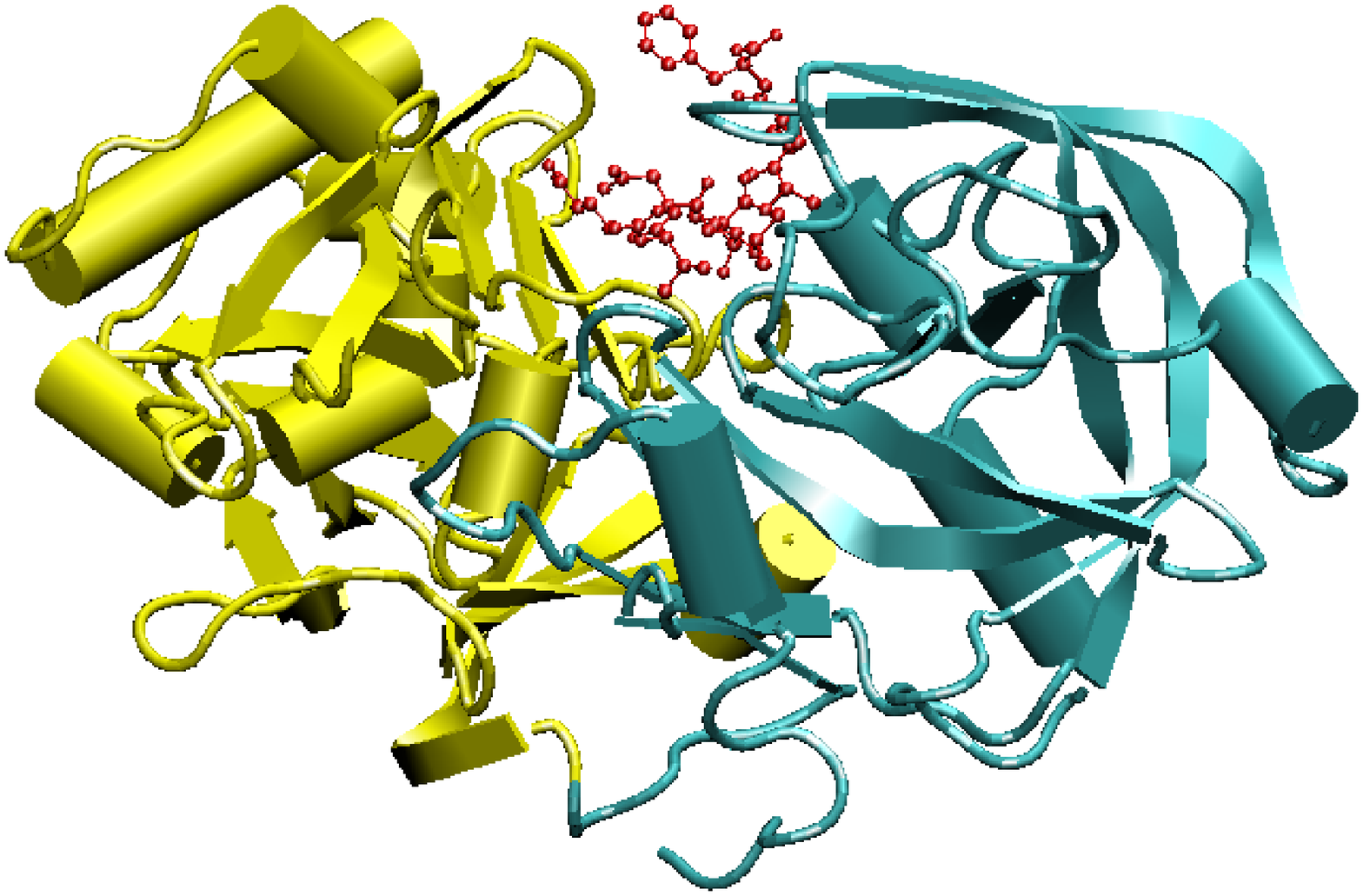} 
(b)\includegraphics[width=2.0in]{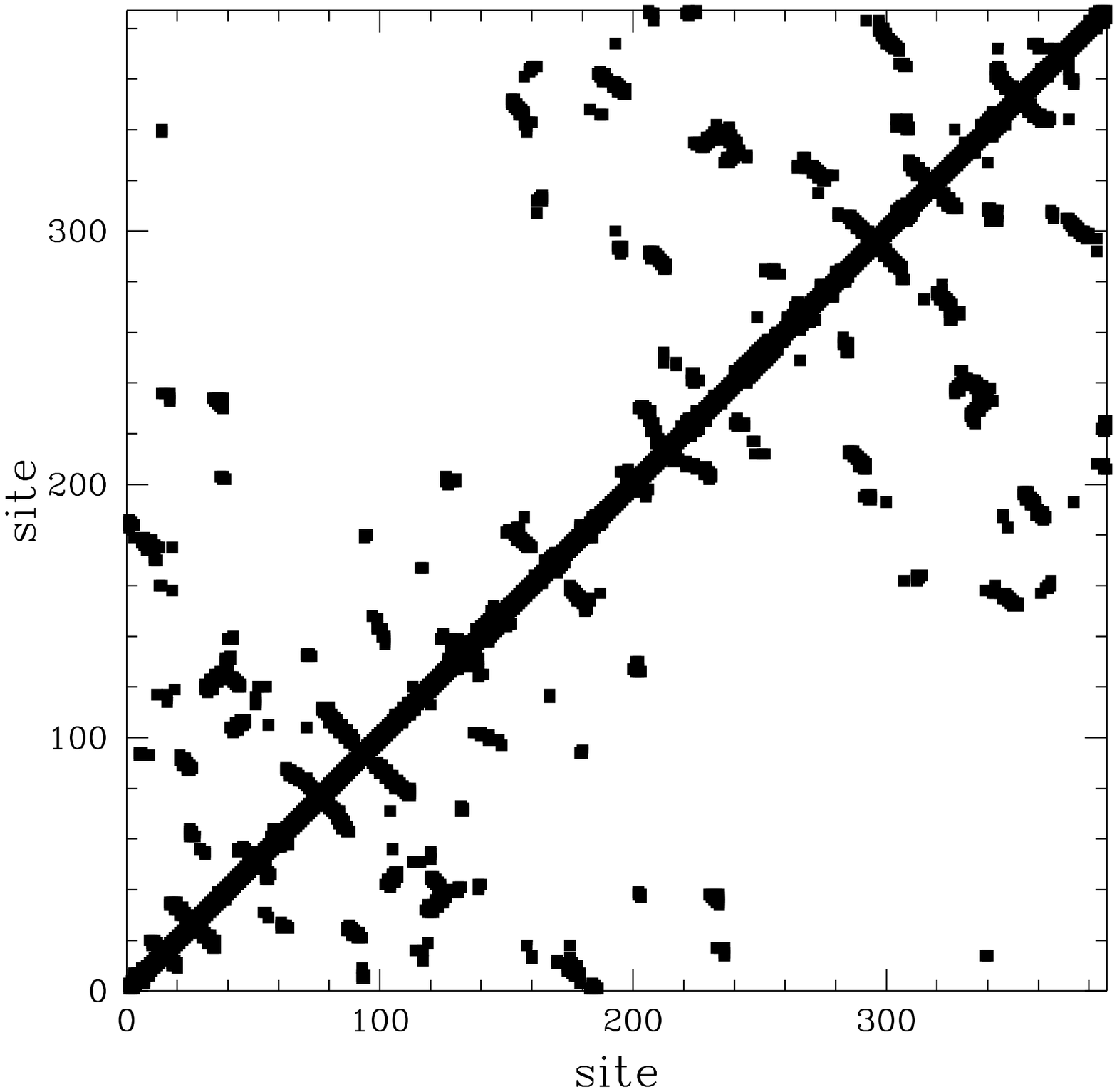}
\caption{{\small (a) Structure of the human $\beta$-secretase (PDB file
1fkn) complexed with an inhibiting peptide (represented with a
wire-frame). The colour highlights the presence of two lobes in the enzyme. 
(b) The contact matrix, $\Delta$, associated to the native (crystallographic)
structure of the enzyme.}}
\label{fig:bace}
\end{center}
\end{figure}

Several studies have attempted to bridge the gap between the time
scales of feasible MD simulations and the ones of
biologically-relevant protein movements by recursing to a mesoscopic
rather than a microscopic approach~\cite{tir96}. A key contribution in
this framework has been the observation that the over-damped dynamics
of a protein in its solvent can be described as occurring in an
effective quadratic
potential~\cite{go82,Karplus85,hinsen98,Karplus76,Karplus82}.  This
observation was further supported by Tirion who pointed out that, in a
normal mode analysis of protein vibrations, the complicated classical
force-field could be replaced by harmonic couplings with the same
spring constants \cite{tir96}. These results stimulated a variety of
studies where the elastic properties of proteins have been described
by means of coarse-grained models where amino acids are replaced by
effective centroids (corresponding to the $C_\alpha$ and/or $C_\beta$
atoms) and the energy function is reduced to harmonic couplings
between pairs of spatially close centroids. These approaches have been
found to be in accord with both experimental and MD characterization
of the overall protein elastic
behaviour~\cite{bah97,bah98,jer99,doruker2000,ani01}.  In particular,
the recently-introduced $\beta$-Gaussian modelhas been shown to be
quite effective in identifying, with a modest computational effort,
the most important conformational changes that a protein can undergo
in thermal equilibrium~\cite{micheletti}. These vibrational modes are
here investigated for a protein, the human $\beta$-secretase (denoted
as BACE hereafter) which is an important representative of the
pepsins, a family of enzymes which are capable of cleaving a peptidic
substrate through a chemical reaction (hydrolysis) involving an
aspartic dyad. It is important to point out that a distinctive feature
of this enzyme is the fact that the two aspartates are located on
opposites sides of the interface between the two lobes. Recently, the
BACE, a membrane-anchored extracellular protein, has been the subject
of several experimental investigations since it constitutes a
key-target for drugs used in treatments of the Alzheimer's disease
(AD)~\cite{Sinha,b-secret}. The onset of this neuro-degenerative
disease is related to the formation of $\beta$-amyloid plaques in
brain tissues.  Recent investigations have demonstrated that the
$\beta$-amyloid peptides are the product of the cleavage of a
precursor protein operated by the BACE. These results have motivated a
large body of scientific investigations aimed at clarifying the
cleavage mechanisms of the BACE and its differences functionality with
respect to typical members of the pepsin family. Interestingly, from a
structural point of view, see Fig.~(\ref{fig:bace}), the core region
of the BACE presents only minor differences from typical pepsin
members. Most of the structural changes are instead located at the
surface of the protein, in the form of six loop insertions, and at the
C-terminus, by a 35-residue long extension\cite{Secret3D}.

As visible in Fig.~(\ref{fig:bace}), where the BACE is shown together
with a bound inhibiting peptide, a long cleft of $35~\mbox{\AA}$ is
present at the interface of the two lobes (highlighted with different
colours). Interestingly, a very small number of residue pairs are
found in contact at the lobes interface and they are mostly
constituted by the aspartic acids involved in the catalytic
activity. In analogy with other
enzymes~\cite{pianaJMB,pianaPsci,micheletti} one may anticipate that
also the BACE catalytic action depend not only on the favourable
chemical interaction of the substrate and the aspartic dyads but also
on the possibility that the conformational fluctuations of the enzyme
itself may module the activity. We shall therefore undertake the task
of characterizing in detail the elastic response of the BACE by
resorting to the $\beta$-Gaussian model. An independent term of
comparison for the robustness of these findings will be provided by an
all-atom Molecular Dynamics simulation. It is found {\em a
posteriori}, that the two methods, which are very different in spirit,
provide a consistent picture for the largest protein rearrangements
occurring in thermal equilibrium. Finally, we characterize the extent
to which the motion of a substrate bound to the enzyme is correlated
with the latter. By these means we identify a limited number of amino
acids in the enzyme which have a strong mechanical bearing of the
conformational fluctuations of the enzyme despite the lack of spatial
proximity.

The programme which implements the $\beta$-Gaussian model is
available, upon request, from the authors.

\section*{The $\beta$-Gaussian Model}\label{BGmodel}

Since the main objective is the modelling of the large scale
fluctuations in a protein it is convenient to reduce the spatial
degrees of freedom of the biopolymer through a coarse-graining where a
two-particle representation is used for each amino acid. Besides the
$C_\alpha$ atom, an effective $C_\beta$ centroid is employed to
capture, in the simplest possible way, the sidechain orientation in a
given amino acid (except for Gly for which only the $C_\alpha$ atom is
retained). This reduced structural representation affects the form of
the effective Hamiltonian associated to the coarse-grained structural
representation. Arguably, the simplest energy function for the system
can be constructed by assuming that all centroids in the protein
($C_\alpha$ and/or $C_\beta$) whose separation is smaller than a given
interaction distance, $R$, interact through the same pairwise
potential, $V$~\cite{tir96}. The information of which centroids are in
interaction in the native state is aptly summarised in the native
contact matrix $\Delta^{XY}_{ij}$ which takes on the values of 1 [0]
if the native separation of the particles of type $X$ and $Y$,
belonging respectively to residues $i$ and $j$, is below [above]
$R$. The contact map of the BACE considered in this study is shown in
Fig.~\ref{fig:bace}. The system energy function evaluated on a trial
structure, $\Gamma$, can then be written as

\begin{equation}
{\cal H} (\Gamma)  = {\cal H}_{BB} (\Gamma) + {\cal H}_{\alpha\alpha} (\Gamma) 
+ {\cal H}_{\alpha\beta} (\Gamma) + {\cal H}_{\beta\beta} (\Gamma) 
\label{eqn:ham}
\end{equation}

\noindent where 
\begin{eqnarray}
{\cal H}_{BB} (\Gamma) &=&  K \sum_i V (d_{i,i+1}^{CA-CA})\nonumber \\
{\cal H}_{\alpha\alpha} (\Gamma) &=& \sum_{i<j} \Delta^{CA-CA}_{ij} V (d_{i,j}^{CA-CA}) \nonumber\\
{\cal H}_{\alpha\beta} (\Gamma) &=& \sum_{i,j} \Delta^{CA-CB}_{ij} V (d_{i,j}^{CA-CB}) \nonumber\\
{\cal H}_{\beta\beta} (\Gamma) &=& \sum_{i<j} \Delta^{CB-CB}_{ij} V (d_{i,j}^{CB-CB}), 
\label{eqn:hamb}
\end{eqnarray}

\noindent In expressions (\ref{eqn:hamb}) we indicated with
$d^{XY}_{ij}$ the actual separation of the particles in the trial
structure, $\Gamma$. The indices $i$ and $j$ run over all integer
values ranging from 1 up to the protein length, $N$. To account for
the protein chain connectivity we have introduced in (\ref{eqn:hamb})
a ``backbone'' energy term which mimics the higher strength of the
bond of consecutive amino acids with respect to non-covalent contact
interactions. Consistently with previous studies~\cite{micheletti} we
have set $R$ equal to $7.5$ \AA\, and $K=1$.

The general pairwise Hamiltonian of eqn (\ref{eqn:ham}) is subject to
an important requirement since it must guarantee that the assigned
reference conformation is at the global energy minimum. The customary
way to accomplish this~\cite{bah97,doruker2000,ani01} is to assume
that the potentials $V^{X-Y}$ appearing in (\ref{eqn:hamb}) attain
their global minimum in correspondence of the native separation of the
centroids. For small fluctuations around the native structure, the
interaction energy of two centroids, $i$ and $j$, can then be expanded
in terms of the deviations from the native distance-vector, ${\bf
r}_{ij}$. If we indicate the deviation vector as ${\bf x}_{ij}$, so
that the total distance vector is ${\bf d}_{ij} = {\bf r}_{ij} + {\bf
x}_{ij}$, we can approximate the pairwise interaction as

\begin{equation}\label{potential}
V({d}_{ij})\approx V({r}_{ij})+\frac{k}{2}\sum_{\mu,\nu}
\frac{r^\mu_{ij}r^\nu_{ij}}{r^2_{ij}}x^\mu_{ij}x^\nu_{ij}
\end{equation}

\noindent where $\mu$ and $\nu$ denote the Cartesian components, $x$,
$y$ and $z$, and $k$ is the second derivative of $V$ at its minimum.
The expansion of Eq.~(\ref{potential}) brings about a dramatic
simplification of Hamiltonian~(\ref{eqn:ham}) which, in fact acquires
a quadratic dependence in terms of the deviation vectors, ${\bf x}$.
However, a further simplification of the energy function can be
achieved by exploiting the fact that the coordinates of the $C_\alpha$
centroids encode, in an almost unique way, the positions of all other
atoms in the protein, and hence also of the $C_\beta$
atoms~\cite{michelettiJPC,michelettiProt04}. The construction scheme
adopted in the $\beta$-Gaussian model (denoted as $\beta$-GM in the
following) defines the location of the $i$-th $C_\beta$ through a
co-planar version of the Park and Levitt construction rule
\cite{cb_construct}:

\begin{equation}\label{cbetas}
{\bf r}_{CB}(i) = {\bf r}_{CA}(i) + l {2 \, {\bf r}_{CA} (i) -
{\bf r}_{CA} (i+1) -{\bf r}_{CA} (i-1) \over | 2 \, {\bf r}_{CA} (i) -
{\bf r}_{CA} (i+1) -{\bf r}_{CA} (i-1)|}
\end{equation}

\noindent where $l= 3$ \AA. Thus, the degrees of freedom can be
reduced to only those of the $C_\alpha$'s. The linear relationship
between the $C_\alpha$ and $C_\beta$ coordinates allows to recast the
Hamiltonian as follows,

\begin{equation}\label{hamiltonian}
{\cal H} = \frac{1}{2} k\sum_{ij,\mu\nu}x_{i,\mu}{\cal
  M}_{ij,\mu\nu}x_{j,\nu}\;, 
\end{equation}
\noindent where $x_{i,\mu}$ is the deviation of $i$-th $C_\alpha$
along the $\mu$ axis and ${\cal M}$ is a $3N$x$3N$ symmetric
matrix. The elastic response of the system is uniquely dictated by the
eigenvalues and eigenvectors of ${\cal M}$.

In thermal equilibrium, each amino acid moves under the action of the
quadratic Hamiltonian (\ref{hamiltonian}) subject to a viscous
friction originating from interactions with the surrounding solvent as
well as with the rest of the protein\cite{hinsen}. The viscous
hindrance of the motion is so important that the protein dynamics
becomes severely
over-damped~\cite{Karplus85,hinsen98,Karplus76,Karplus82,hinsen}. In
this case it seems appropriate to describe the dynamics of the amino
acids within a Langevin framework~\cite{howard,doi}

\begin{equation}\label{langevin}
\gamma_i~\dot{x}_{i,\mu}(t)= - k\sum_{j,\nu}{\cal
    M}_{ij,\mu\nu}x_{j,\nu}(t)+ \eta_{i,\mu}(t)\;,
\end{equation}

\noindent where $\gamma_i$ is the viscous friction coefficient and
$\eta_{i,\mu}(t)$ is a stochastic noise satisfying the 
relations~\cite{doi,howard}
\begin{equation}
\left\langle~\eta_{i,\mu}(t)~\right\rangle = 0 \;,\;
\left\langle~\eta_{i,\mu}(t)\eta_{j,\nu}(t')~\right\rangle = 
\delta_{ij}~\delta_{\mu\nu}~\delta(t-t')~2k_BT~\gamma_i\;. 
\end{equation}

Within this dynamical framework it is possible to calculate exactly
how correlations among the displacements of various pairs of residues
decay as a function of time. We start by considering the case where
the various viscous coefficients in eqn (\ref{langevin}) taken on
the same value, $\gamma$. In this case one has~\cite{chandrasekhar}:
\begin{equation}\label{correlation1}
\left\langle x_{i,\mu}(t)x_{j,\nu}(t+\Delta t)\right\rangle~
=~\frac{K_BT}{k}\sum_l v_i^l v_j^l\frac{1}{\lambda_l}
e^{-\lambda_l\frac{k}{\gamma}\Delta t}\;,
\end{equation}

\noindent where the average $\left\langle~.~\right\rangle$ in eqn
(\ref{correlation1}) denotes the usual canonical thermodynamics
average. The vector ${\bf v}^l$ and the scalar $\lambda_l$ are,
respectively, the $l$-th eigenvector and the $l$-th eigenvalue of the
matrix ${\cal M}$ ordered in such way that
$\lambda_l~<~\lambda_{l+1}$. It should be noted that since the
Hamiltonian is invariant under rotations and translations, it will
always possess (at least) six eigenvectors associated to eigenvalues
equal to zero that are obviously excluded from the sum in expression
(\ref{correlation1}) and related ones~\cite{micheletti,hiv-gauss}. The
eigenvectors $\{ {\bf v}^l\}$ represent therefore the independent
modes of structural relaxation in the protein, while the associated
decay times are given by
\begin{equation}
\tau_l~=~\frac{\gamma}{k\lambda_l}\;.
\label{eqn:decaytimes}
\end{equation}
\noindent Of particular interest, for the purpose of identifying the
concerted large scale structural fluctuations occurring in a protein,
is the degree of correlation of pairs of residues at equal times. This
information is summarised in the covariance matrix, ${\cal C}$, whose
elements are obtained by $\Delta t =0$ in (\ref{correlation1})

\begin{equation}\label{correlation2}
{\cal C}_{ij,\mu\nu} \equiv \left\langle x_{i,\mu}x_{j,\nu}\right\rangle=
             \frac{K_BT}{k} \sum_l {v^l_i v^l_j \over \lambda_l} =
             \frac{K_BT}{k}~{\cal M}^{-1}_{i,j,\mu,\nu}\;.
\end{equation}

\noindent Due to the fact that this matrix contains the full
three-dimensional information about pair correlations its linear size
is $3N$. Typically it is important only to quantify the relative
degree of correlation of any two residues; in this case one can
consider the normalised reduced covariance matrix (of linear size $N$)
which is defined as

\begin{equation}\label{eqn:redcovmat}
{\cal C}_{ij} = \frac{\left\langle{\bf x}_i \cdot {\bf
x}_j\right\rangle} {\sqrt{\left\langle~|\bf{x}_i|^2\right\rangle
\left\langle~|\bf{x}_j|^2\right\rangle}} = {{\cal C}_{ij,\nu\nu} \over
\sqrt {{\cal C}_{ii,\nu\nu} {\cal C}_{jj,\mu\mu}}}
\end{equation}

\noindent where a sum is implied over repeated indices. It is
important to stress that the elements of the covariance matrix are
thermodynamic averages reflecting equilibrium properties of the system
and, hence, are independent of the friction coefficients. Through
eqn~(\ref{correlation2}) one sees that the same set of vectors $\{
{\bf v}^l\}$ describing the modes of relaxation also describes the
independent modes of structural fluctuations in thermal
equilibrium. The contribution of each mode to the pair correlation is
inversely proportional to $\lambda_l$ thus establishing the intuitive
result that the modes associated to the longest relaxation times are
those responsible for the largest structural fluctuations. The
equivalence of the vectors describing the modes of relaxation and
structural fluctuations is a consequence of having used the same
viscous coefficient for all amino acids in the protein.  When this
simplifying assumption is not made one has that the modes and times of
relaxation are determined by diagonalizing instead the symmetric
matrix

\begin{equation}\label{symmetric}
\widetilde {\cal M}_{i,j,\mu,\nu}= \frac{{\cal
M}_{i,j,\mu,\nu}}{\sqrt{\gamma_i\gamma_j}}\;.
\end{equation}

\noindent On the other hand, the modes of structural fluctuation are
the same as before, as can be ascertained by calculating directly the
thermodynamic average $\langle x_{i,\mu}x_{j,\nu}\rangle$ with the
canonical weight associated to Hamiltonian (\ref{hamiltonian}).

\section{Conformational fluctuations of the BACE}

\begin{figure}[htbp]
\begin{center}
\includegraphics[height=5cm]{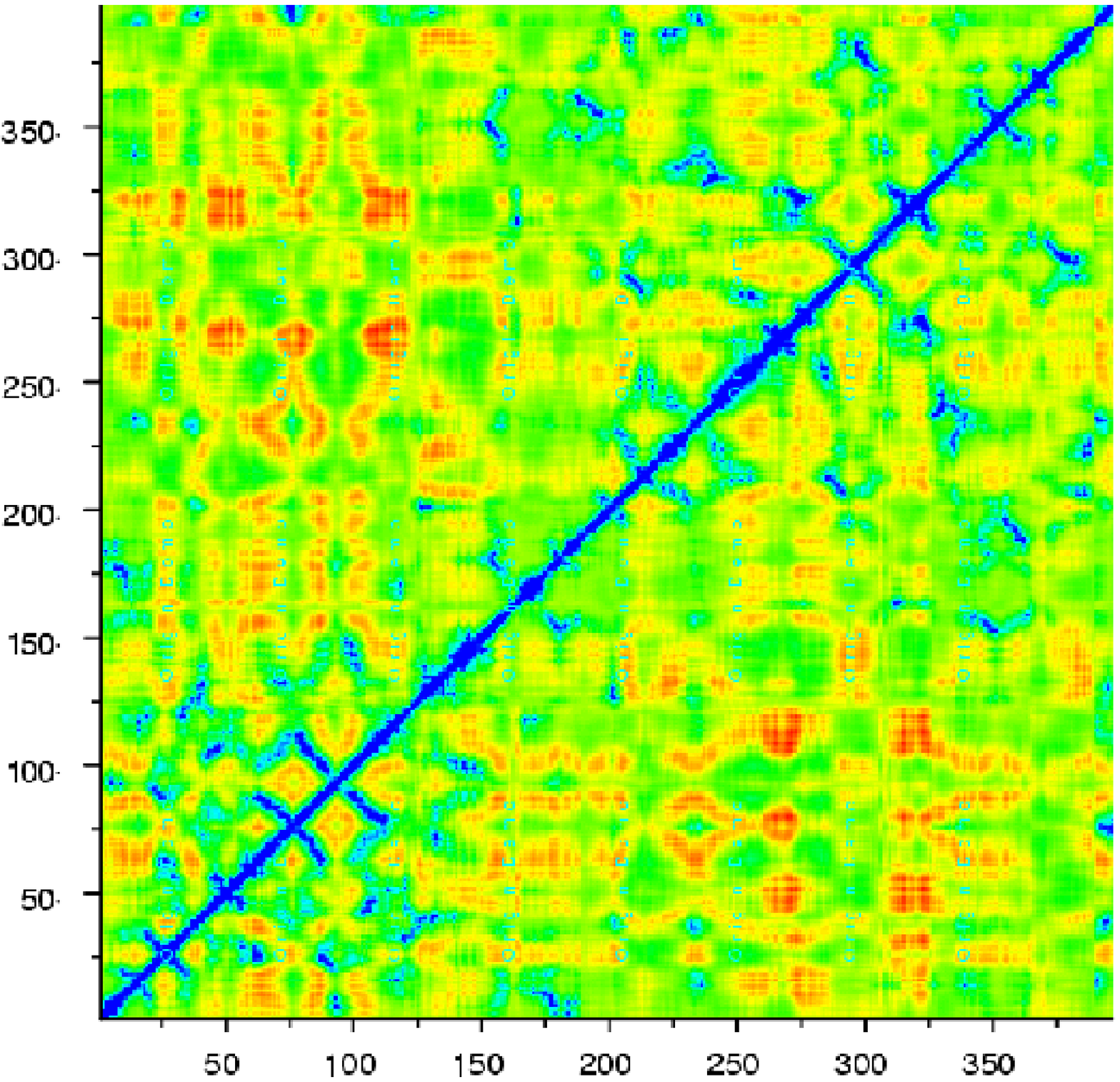}
\caption{{\small Reduced covariance matrices for BACE, as obtained from
           $\beta$-Gaussian model computations. Entries with values close to 1
(strong positive correlation) are shown in blue, while anti-correlated
pairs are shown in red.}}
\label{covmats}
\end{center}
\end{figure}

We have applied the above-mentioned analysis to the native structure
of the BACE shown in Fig.~(\ref{fig:bace}a). The associated normalised
reduced covariance, computed through the $\beta$-GM is shown in
Fig.~(\ref{covmats}). The comparison of the covariance matrix with the
contact map of Fig.~(\ref{fig:bace}b) shows that the highest positive
correlations are observed, as expected, in correspondence of
contacting residues. More interesting is the presence of negative
correlations which signal important mechanical couplings between
regions that are not in spatial
proximity~\cite{pianaJMB,pianaPsci,micheletti}.  The inspection of
Fig.~(\ref{covmats}) reveals a significant degree of anti-correlation
among residues on one lobe (especially 40-55 and 100-114) and those on
the other one (especially 260-272 and 305-320). The fact that the
conformational fluctuations of the two lobes are, on average, directed
in opposite directions suggests that the enzymatic functional modes
may be based on an opening/closing mechanism of the lobes.

\begin{figure}
\begin{center}
\includegraphics[width=7cm]{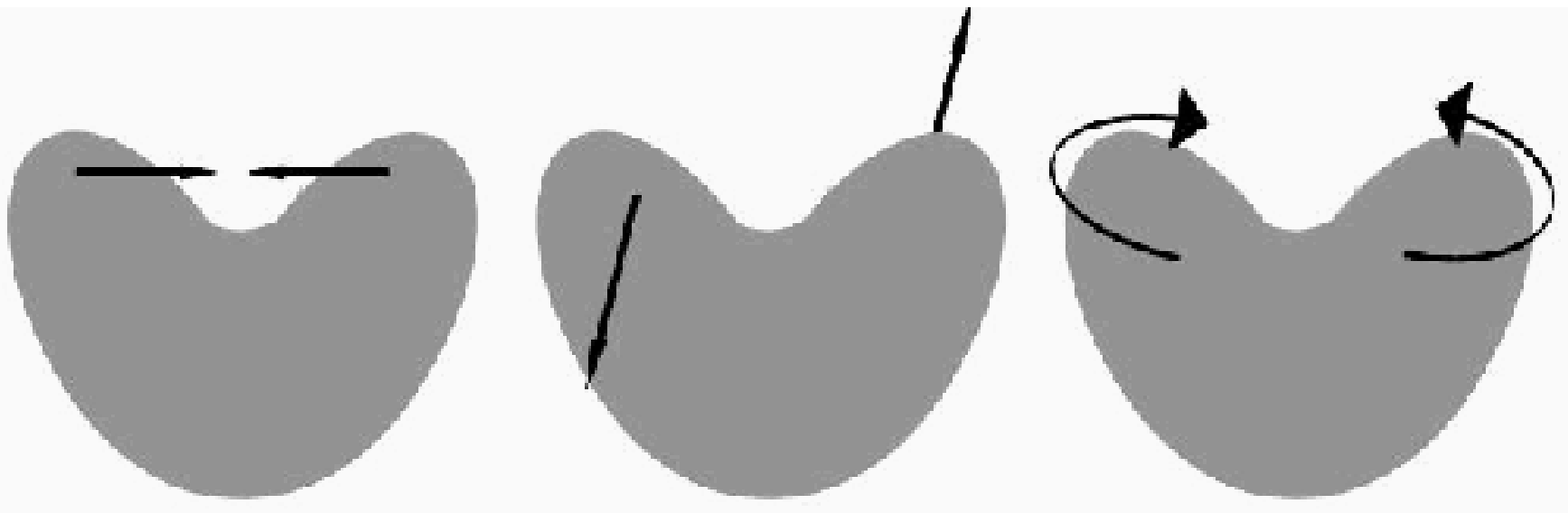}
\caption{{\small Pictorial representation of the first three modes
    (from left to right) which describe the largest conformational
    distortions in the BACE.}}
\label{modi}
\end{center}
\end{figure}

This insight can be considerably refined by examining the individual
contribution of the various eigenvectors of ${\cal M}$ to the pair
correlations. It is apparent from the decomposition of
Eq.~(\ref{correlation2}) that the weight of the eigenvectors is
inversely proportional to the corresponding eigenvalues. For this
reasons we shall now describe the structural deformations encoded by
the first three eigenvectors which, alone, are responsible for a good
fraction of the overall residue mobility. By superimposing to the
native structure the distortion associated to the first mode one can
ascertain that it involves the movement of the lobes along opposite
directions with respect to the plane identified by the cleft of the
enzyme (each lobe moving almost rigidly while the connecting regions
between the lobes is almost static). In the second mode the
anti-correlated motion of the lobes is still visible but this time the
motion occurs mostly parallel to the cleft plane, resulting in a shear
deformation, as depicted in Fig.~(\ref{modi}) while in the third mode,
the two lobes appear to rotate in counter-directions. In all three
modes the high mobility of the exposed loop ranging from Val$309$ to
Asp$317$ is very noticeable.

So far we have summarised the results obtained with the $\beta$-GM
which relies on several simplifying assumptions (including the use of
the same friction coefficient for all residues, a limitation that will
be removed later). It is therefore important to verify the validity of
the conclusions reached here against independent terms of
reference. Ideally, one would like to compare the model results
against direct experimental determination of the quantities of
interest here, such as the correlation of residues' motion. However,
this is not presently feasible in a direct way and hence such detailed
information can only be obtained from dynamical simulations with
all-atom interaction potentials.  We therefore carried out a MD
simulation of the BACE in explicit solvent. The simulated system was
constituted by the protein immersed in a water box of size $75$~\AA\
~x~$87$~\AA\ ~x~$90$~\AA\ to which $9$ sodium counter-ions were added
to ensure the overall charge neutrality. The whole system, composed of
about $48,000$ atoms, underwent $20$~ns of MD simulations (after
relaxation) carried out at 300 K with the GROMACS
program~\cite{GROMACSa,GROMACSb}. The Amber force-field
(Parm98~\cite{AMBER}) was used to describe the protein and the
counter-ions, whilst the TIP3P model was used for water. Particle mesh
Ewald routines were used to treat long-range electrostatic
interactions. A cut-off of 12~\AA\ was used for the Van der Waals
interactions and the real part of the electrostatic interactions.

We carried out the comparison of the $\beta$-GM and molecular dynamics
through a series of steps which include the comparison of the overall
mobility of the various amino acids, of corresponding entries of the
covariance matrices and of the largest eigenvectors of the covariance
matrix. These comparisons are here exploited not only to check the
consistency of the two approaches, but also as a means to assign the
$\beta$-GM parameter $k$ from the MD comparison.

The first quantity that we shall compare is the mean-square
fluctuation which summarises the overall mobility of any given
residue. Within the $\beta$-GM this quantity is given by the diagonal
elements of ${\cal M}^{-1}$

\begin{equation}
\left\langle{\bf x}_i \cdot {\bf x}_i\right\rangle = \sum_{\alpha}
{\cal M}^{-1}_{ii,\alpha,\alpha}\;,
\label{eqn:mobility}
\end{equation}

\noindent while, in the context of Molecular Dynamics, the
thermodynamic average of Eq.~(\ref{eqn:mobility}) is aptly replaced by
the time average over the simulation run. It should also be noted
that, usually, it is not easy to ascertain whether the simulated
trajectory is sufficiently long that thermodynamic averages can be
legitimately replaced with dynamical ones. A practical way of checking
{\em a posteriori} the validity of this ergodicity assumption, is to
check the consistency of the essential eigenspaces pertaining to
different parts of the simulated trajectories. Indeed, this type of
analysis shows that when the simulated time span greatly exceeds the
typical time-scale associated to the slow dynamical modes then the
essential subspaces are robustly identified\cite{Amadei99}. Besides
this strategy, a recent theoretical study has introduced some valuable
quantitative criteria by which it is possible to decide whether a
given MD trajectory is too short to meet the ergodicity
requirements\cite{hessb}. These criteria are, again based on the
analysis of the essential dynamical spaces. When the dynamical
sampling of phase-space is insufficient then the eigenvectors of the
essential spaces have distinctive cosine-like shapes. By converse, it
has been shown that when the cosine content of the relevant dynamical
eigenspaces is negligible then the ergodicity assumption appears
justified\cite{hessb}. In the present study we have adopted both these
criteria: the total simulation time span of 20 ns is about thirty
times bigger than the longest autocorrelation time found in the system
(see last section); in addition the cosine content of the top 5
essential eigenvectors was, on average, below 10 \%.

\begin{figure}[htbp]
\begin{center}
\includegraphics[width=7cm]{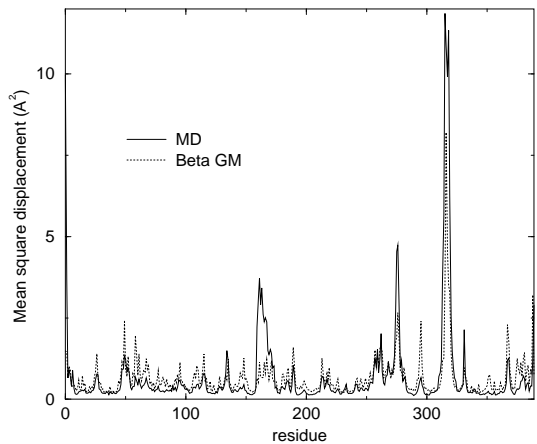}
\caption{{\small Mean square fluctuations of the various residues in
    the BACE obtained from MD and calculated from the
    $\beta$-GM after an optimal choice of the harmonic coupling
    parameter, $k$.}}
\label{b-value}
\end{center}
\end{figure}

It must be noticed that, within the $\beta$-GM the mean square
displacements are proportional to $k_B T / k$. The comparison of the
MD and model mean square displacements offers, therefore, the
opportunity to set the effective value of $k$ by, e.g.  matching the
average mean square displacements in the two cases. This criterion
yields, for this particular protein, the value $k^{-1} = 0.7$ \AA$^2 /
K_B\,T$.  The profiles for the residues mobility according to MD and
the $\beta$-GM (having fixed $k$ to the value mentioned above) are
shown in Fig.~(\ref{b-value}).

As mentioned before, Fig.~(\ref{b-value}) reveals the high mobility of
the exposed loop spanning residues $309$-$317$. Aside from this, the two
profiles have a good degree of correlation since the linear
correlation coefficient is $0.78$ while the more stringent Kendall
(non-parametric) coefficient one is $0.61$~\cite{halle2002,NR}.

Having ascertained the consistency of the overall residues mobility of
MD and $\beta$-GM we analyze the agreement of the covariance matrices
which convey information on pair correlations. We carry out the
comparison in two stages: first by comparing corresponding entries of
the matrices and subsequently by measuring the overlap between the
significant subspaces of the matrices.

\begin{figure}[htbp]
\begin{center}
\includegraphics[width=7cm]{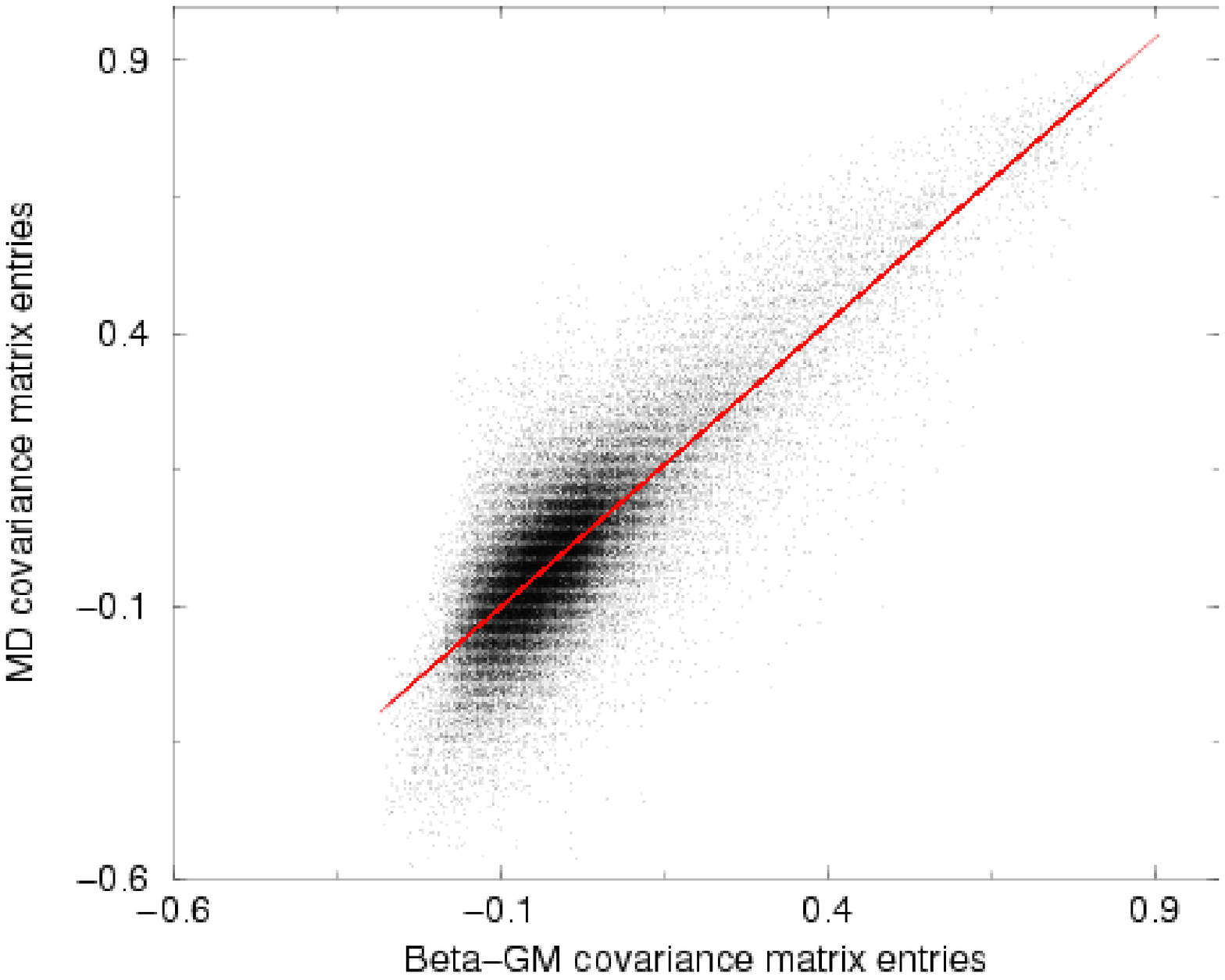}
\caption{{\small Scatter plot of corresponding entries of the reduced
    covariance matrices obtained by the $\beta$-Gaussian model and MD
    simulations of BACE. The interpolating line, determined
    by linear regression, is also shown.}}
\label{scatterplot}
\end{center}
\end{figure}

The degree of accord of the normalized reduced covariance matrices of
the MD and $\beta$-GM is conveniently summarised through a scatter
plot of corresponding entries of the two matrices. The results are
summarised in Fig.~(\ref{scatterplot}) where we have deliberately
omitted the diagonal entries of the matrix which are equal to 1 in
both cases. Since the length of the BACE approaches $400$ residues,
the data set shown in Fig.~(\ref{scatterplot}) is constituted by more
than $7.5~10^4$ distinct data points. The linear correlation
coefficient of the MD and $\beta$-GM data is $r=0.83$.  This represent
a strikingly high value given the huge number of degrees of
freedom. However, a precise quantification of its statistical
significance cannot be conveyed since the data are manifestly not
distributed in a binormal fashion and the Kendall non-parametric
analysis is inapplicable to to the excessive number of data
points~\cite{micheletti}. We conclude the discussion of the scatter
plot of Fig.~(\ref{scatterplot}) by mentioning that, in case of
perfect correlation of two normalised (adimensional) covariance
matrices the data would align along the diagonal of the graph in
Fig.~(\ref{scatterplot}). Interestingly, the best-fit line lies very
close to the diagonal having a slope of $s=1.04$, a fact which further
testifies the overall viability of the Gaussian scheme.

As a final test we wish to measure the consistency of the most
important eigenspaces of the two matrices, that is the eigenvectors
associated to the largest eigenvalues of the covariance matrix (for
$\beta$-GM this is equivalent to considering the {\em smallest}
eigenvalues of ${\cal M}$). In fact, these eigenspaces are the ones
that describe the most significant modes of distortion of the molecule
in the three-dimensional space. Various strategies have been proposed
to evaluate the accord of the significant eigenspaces of the
covariance matrices~\cite{hinsen98,covmat,hessb}. Here we have adopted
the simple strategy of considering the top 10 eigenvectors obtained
from MD and the $\beta$-GM and calculating the modulus of the scalar
product of any pair of vectors coming from the two sets. It is found
that the overlap between the two essential eigenspace is considerable
since about 75 \% the norm of the first four eigenvectors of the
$\beta$-GM are projected on the first four eigenvector of MD. Overall
the root mean square inner product suggested by Amadei {\em et al.}
\cite{Amadei99} among the two sets was 0.6 which denotes a non trivial
correlation, although a precise assessment of its statistical
significance is difficult to ascertain \cite{Amadei99,hessb}.

\begin{figure}[t]
\begin{center}
\includegraphics[width=7cm]{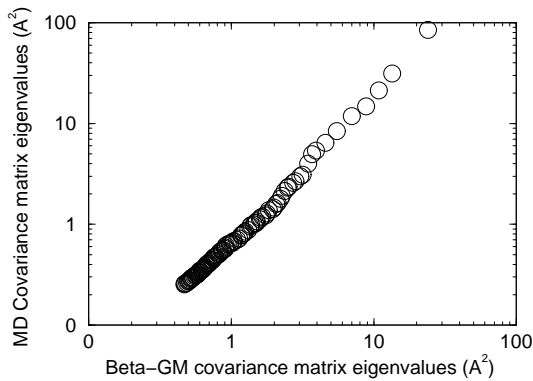}
\caption{{\small Scatter plot of equally ranking eigenvalues of the
    MD and $\beta$-GM covariance matrix.}}
\label{eigenvalues}
\end{center}
\end{figure}

It is also interesting to evaluate the agreement of corresponding
(i.e. equally ranking) eigenvalues of the covariance matrices. This
analysis is carried out in the scatter plot of
Fig.~(\ref{eigenvalues}) where, for the $\beta$-GM the $i$th largest
ranking eigenvalue of the covariance matrix was taken as the inverse
of the $i$th smallest eigenvalue of ${\cal M}^{-1}$ multiplied by the
previously found coefficient $k_B T / k$. Since the eigenvalues
span a few orders of magnitude, the plot is presented in a log-log
format and reveals a relationship that rather than being simply linear
is better described by a power-law dependence with exponent of about
$1.4$.

\begin{figure}[t]
\begin{center}
\includegraphics[width=7cm]{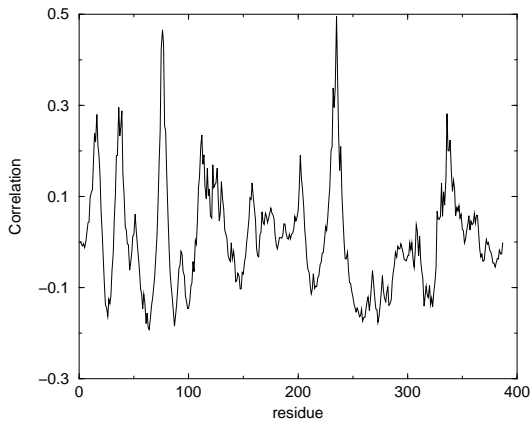}
\caption{{\small The curve indicates the degree of correlation,
    calculated with $\beta$-GM, of a residue in the middle of the
    inhibiting peptide and the 387 residues that constitute the BACE.}}
\label{substrate}
\end{center}
\end{figure}

After having characterize the accord of the conformational
fluctuations occurring in thermal equilibrium within the all-atom MD
and the $\beta$-GM we focus on the characterization of the mechanical
coupling between the BACE enzyme and the substrate (peptide) on which
it acts. This analysis is accomplished by applying the Gaussian model
not to the isolated enzyme but the enzyme/substrate complex shown in
Fig.~(\ref{fig:bace}). The crucial questions to formulate in this
context regard the possible existence of non-trivial mechanical
couplings between the substrate and residues in the enzyme. For other
enzymes, in particular the HIV-1 protease, it has been shown that
several residues of HIV-1 protease despite being far away from the
substrate have an important mechanical bearing on the
latter~\cite{micheletti,pianaJMB}.  It is by virtue of this mechanical
coupling that it is possible to rationalize the emergence of mutations
causing drug resistance in correspondence of sites far from the
cleavage region (e.g. Met63Ile, Met46Ile-Leu and
Leu47Val)\cite{condra1,boucher1,Molla}.  In fact, as the explicit MD
calculation has shown, the high degree of such coupling between such
sites and the cleavage is such that the detailed chemical identity of
the former strongly influence the substrate binding affinity of the
latter\cite{pianaPsci}.  Therefore, mechanical couplings can be so
important for enzymatic catalysis that mutations at a small number of
key enzymatic sites (even if distant from the active site) can
dramatically alter the enzyme reactivity. We have therefore undertaken
a similar analysis here and calculated the degree of correlation of
the displacement of residues in the substrate and all residues in the
BACE. The correlation profiles are nearly identical for all residues
in the substrate and a representative profile is shown in
Fig.~(\ref{substrate}).

From the profile a strong positive correlations is discernible in
correspondence of residues $72$ and $230$ (numbering according to the
PDB file). This reflects the close spatial proximity of the substrate
and these contacting residues. The crucial feature is however
contained in the peaks of negative correlations which occur in
correspondence of residues $23$, $61$, $83$, $255$, $269$,
$317$. These residues are very distant from the substrate ($25$ - $30$
\AA) and yet are able to influence the conformational fluctuations of
the substrate bound to the active site. These sites are located at the
outward face of each of the two lobes, in close analogy to the key
mechanical sites found for HIV-1 protease. This fact provides
additional evidence in support of the fact that the steering of the
substrate towards a reactive conformation is achieved not only through
finely-tuned chemical couplings with the enzymes but also through
subtle mechanical influence of sites far apart from the active region.

\begin{figure}
\begin{center}
\includegraphics[width=7cm]{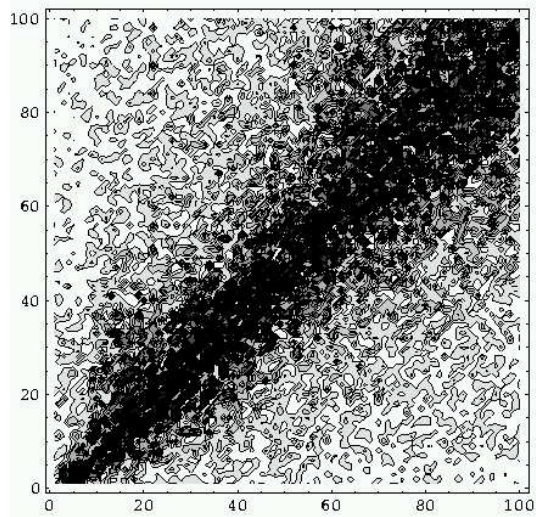}
\caption{{\small Density plot for the modulus of the scalar product of
    the eigenvectors of matrix ${\cal M}$ (rank index given on $x$ axis)
    against those of matrix $\tilde{\cal M}$ (rank index given on $y$
    axis). Entries with values close to 1 [0] are shown in black [white].}}
\label{gamma_i}
\end{center}
\end{figure}

We conclude our analysis by discussing the possibility to gain insight
through the $\beta$-GM on the time scales involved in the most
important conformational fluctuations. To do so it is necessary to
have a quantitative estimate for the effective friction coefficients
appearing in Eq.~(\ref{langevin}). It has been pointed out
before~\cite{karplus_ACR,hinsen98} that the effective friction
coefficient experienced by the various amino acids in a protein depend
not only on the interaction with the solvent but are severely affected
by protein density itself in their neighbourhood. In particular,
Hinsen {\em et al}~\cite{hinsen} have recently provided a
phenomenological characterization of the viscous coefficient which
they found to have an approximate linear relationship in terms of the
local atomic density. We have translated this phenomenological
relationship in terms of the number of contacts, $n^c_i$ to which
residue $i$ takes part (calculated with a cutoff distance of $7.5$
\AA) and thus arrived to the phenomenological relationship

\begin{equation}
\gamma_i = B \, n^c_i
\end{equation}

\noindent where $B = 1.53 \, 10^3$ amu/ps.  We have therefore
calculated the various modes relaxation in this new context where the
friction coefficient is not the same for all residues. Strikingly we
have found that that there is a good correspondence between the modes
of relaxation (which depend on $\gamma_i$ and reflect dynamical
properties) with the eigenvectors which describe conformational
fluctuations in thermal equilibrium. This correspondence is obviously
exact in the case when $\gamma_i$ is the same for all sites but is not
otherwise expected in case of heterogeneity of the friction
coefficient. The degree of accord of the two sets of eigenvectors is
shown in Fig.~(\ref{gamma_i}) and indeed the thick cloud around the
diagonal highlights their good correspondence. The time scales which
controls the decay of conformational fluctuations in a dynamical
trajectory are obtained from the inverse of the eigenvalues of
$\widetilde{\cal M}$ through eq. \ref{eqn:decaytimes}. Therefore, we
obtain that the slowest relaxation time calculated within the Gaussian
model has a decay time of 1.2 ns.  The analysis of the decay of
autocorrelation in the whole MD trajectory indicates 0.7 ns as the
slowest relaxation time of the system. It appears therefore, that the
simple Gaussian approach can correctly identify the time scales of the
system autocorrelation within a factor of two. Therefore, despite the
several approximations which are at the basis of the $\beta$-GM the
latter appears to be useful also for estimating, with a modest
computational expenditure, the correct order of magnitude of the
system autocorrelation time. This fact may be exploited to have a
preliminary indication of the lapse of time that needs to be covered
in an all-atom MD simulation to ensure that the system has sufficient
time to explore significantly different regions of the phase space.

\section{Conclusions}

We have studied the conformational fluctuations of BACE with the
purpose of clarifying the effect of conformational fluctuations on the
capability of enzymes to steer the substrate in a favourable reactive
conformations. In analogy with results obtained on a different class
of enzymes ~\cite{pianaJMB,pianaPsci,micheletti} also for this enzyme
it is found that a few sites that are located remotely from the active
site have a strong mechanical influence on the enzyme-substrate
couplings. The analysis was carried out using a simple and
computationally inexpensive Gaussian network scheme. Several aspects
of the results obtained within the simplified coarse-grained model
have been validated and tested against an extensive all-atom MD
simulation.

\subsection*{Acknowledgements}

We are indebted to Paolo Carloni and Luca Marsella for several
illuminating discussions and suggestions. We acknowledge support from
INFM-Democritos and MIUR Cofin 2003.


\end{document}